\documentclass[reprint,aps,prx,amsmath,amssymb,superscriptaddress,floatfix,longbibliography]{revtex4-2}
\usepackage{amstext}
\usepackage{graphicx}
\usepackage{xcolor}

\begin{document}
\title{Reaching Quantum Critical Point by Adding Non-magnetic Disorder\\ in Single Crystals of Superconductor $(\text{Ca}_x\text{Sr}_{1-x})_3\text{Rh}_4\text{Sn}_{13}$}

\author{Elizabeth~H.~Krenkel}
\affiliation{Ames National Laboratory, Ames, Iowa 50011, U.S.A.}
\affiliation{Department of Physics \& Astronomy, Iowa State University, Ames, Iowa
50011, U.S.A.}

\author{Makariy~A.~Tanatar}
\affiliation{Ames National Laboratory, Ames, Iowa 50011, U.S.A.}
\affiliation{Department of Physics \& Astronomy, Iowa State University, Ames, Iowa 50011, U.S.A.}

\author{Romain~Grasset}
\affiliation{Laboratoire des Solides Irradi\'{e}s, CEA/DRF/lRAMIS, \'{E}cole Polytechnique, CNRS, Institut Polytechnique de Paris, F-91128 Palaiseau, France}

\author{Marcin~Ko\'{n}czykowski}
\affiliation{Laboratoire des Solides Irradi\'{e}s, CEA/DRF/lRAMIS, \'{E}cole Polytechnique, CNRS, Institut Polytechnique de Paris, F-91128 Palaiseau, France}

\author{Shuzhang~Chen}
\affiliation{Condensed Matter Physics and Materials Science Department, Brookhaven
National Laboratory, Upton, New York 11973, U.S.A.}
\affiliation{Department of Physics and Astronomy, Stony Brook University, Stony
Brook, New York 11794-3800, U.S.A.}

\author{Cedomir~Petrovic}
\affiliation{Condensed Matter Physics and Materials Science Department, Brookhaven
National Laboratory, Upton, New York 11973, U.S.A.}
\affiliation{Department of Physics and Astronomy, Stony Brook University, Stony
Brook, New York 11794-3800, U.S.A.}
\affiliation{Shanghai Key Laboratory of Material Frontiers Research in Extreme Environments (MFree), Shanghai Advanced Research in Physical Sciences (SHARPS), Pudong, Shanghai 201203, China}

\affiliation{Department of Nuclear and Plasma Physics, Vinca Institute of Nuclear Sciences, University of Belgrade, Belgrade 11001, Serbia}

\author{Alex~Levchenko}
\affiliation{Department of Physics, University of Wisconsin--Madison, Madison,
Wisconsin 53706, U.S.A.}

\author{Ruslan~Prozorov}
\email[Corresponding author: ]{prozorov@ameslab.gov}
\affiliation{Ames National Laboratory, Ames, Iowa 50011, U.S.A.}
\affiliation{Department of Physics \& Astronomy, Iowa State University, Ames, Iowa 50011, U.S.A.}

\date{18 March 2026}

\begin{abstract}
The Remeika series superconductor, $(\text{Ca}_x\text{Sr}_{1-x})_3\text{Rh}_4\text{Sn}_{13}$, shows a rare nonmagnetic quantum critical point (QCP) associated with the continuous charge-density wave (CDW) and structural transition under the ``dome'' of superconductivity achieved by tuning composition and applying pressure. Here we use a nonmagnetic point-like disorder induced by 2.5 MeV electron irradiation to suppress the CDW and drive the system to and even beyond the QCP. This conclusion is based on a clear evolution of temperature-dependent resistivity, $\rho\left(T\right)$, from the Fermi liquid to the non-Fermi liquid regime with increasing amount of disorder. Starting on the CDW side, below the suggested QCP concentration of $x_c=0.9$, added disorder resulted in a progressively larger linear term and a reduced quadratic term in $\rho\left(T\right)$. Nearly perfect $T-$linear dependence is observed at the dose at which long-range CDW order is suppressed to $T=$0, consistent with the expectations.
We refine the QCP location in this system and place it in the interval between $x=$0.75 and 0.85.
Our results strongly support the concept that the disorder can tune the system to the quantum critical regime and even beyond. It follows from the argument by Imry and Ma that any ordered phase is unstable toward quenched disorder. Introduced in a controlled way, this disorder becomes a novel non-thermal tuning parameter likely applicable to a variety of different systems.
\end{abstract}

\maketitle
\section*{Introduction}
Phase transitions are associated with broken symmetries, such as translational, rotational, time-reversal, or unitary U(1) and SU(2) \cite{vojta2000,sachdev2000,vojta2003,Belitz2005,belitz2005a,Vojta2006,sachdev2008,schakel2008,kinross2014,brando2016,chandra2017,heyl2018,beekman2019,Khodas:PRB20,kam2023,Custers,cui2025}. Conventional phase transitions are driven by thermal fluctuations, with temperature as a control parameter. In contrast, quantum phase transitions (QPTs) occur at absolute zero and are driven by quantum fluctuations. The QPTs are achieved by varying non-thermal parameters, such as magnetic field, composition, pressure, and strain. The strength of quantum fluctuations increases when the temperature is lowered, but the interval of the driving parameter where it happens shrinks, ultimately to a point at $T \to 0$, called the “quantum critical point” (QCP). Quantum criticality can be studied experimentally, because quantum fluctuations affect many physical properties far from the QCP, in some cases at tens of kelvin, and in a wide interval of the tuning parameter \cite{vojta2000,sachdev2000,vojta2003,Belitz2005,Khodas:PRB20,worasaran2021,kam2023,Custers}.

The picture is far more complex when, instead of one long-range ordered state terminating at the QCP, there are two (or more) microscopically coexisting long-range quantum orders
\cite{vojta2000,Belitz2005,Khodas:PRB20,Carvalho:PRB20,basov2011,chubukov2020,chubukov2015,fradkin2015,mukasa2023,cui2025,wang2025d}. A well-studied case is iron-based superconductors (IBS) where the ``dome" of superconductivity develops when the spin-density wave (SDW) is sufficiently suppressed \cite{fernandes2010,fernandes2014,chubukov2015,decarvalho2020,chubukov2012}.
It has been proposed that magnetic quantum fluctuations could act as a bosonic “glue” for Cooper pair formation
\cite{Mathur1998,putzke2014,kinross2014,biswas2015,freitas2016,gruner2017,Chen2019,wang2025d}. However, what happens with the SDW transition itself under the dome is the subject of current studies which show that there is a QCP inside the dome where $T_{SDW}\rightarrow 0$ \cite{Hashimoto2012,Taka2017,wang2025d}. Another possible long-range ordered phase, studied in this work, is the charge-density wave (CDW) coexisting with superconductivity. CDWs are the periodic commensurate or incommensurate modulations of electron density, usually accompanied by lattice distortion \cite{Gruner1994,johannes2008,overhauser1978,Peierls2001}.

Two ordered quantum phases can interact in a non-trivial way, coexisting microscopically through coupling to soft modes (e.g., phonons), resulting in an intricate interplay \cite{Hashimoto2012,Joshi2020,chowdhury2015}. Research shows that superconductivity is enhanced on approaching the QCP \cite{mukasa2023,biswas2015,freitas2016,gruner2017,putzke2014}. In general, there is a significant effort in the community to establish the general trends and properties of multi-phase QCP physics, which could help unlock the mechanisms of unconventional superconductivity
\cite{vojta2000,basov2011,fradkin2015,jost2022,fazio1995,Veiga2020,chubukov2015,johnston2010,norman2008,stewart2011}.

The interpretation of the results obtained from using different tuning parameters is not straightforward due to the simultaneous effect on the electronic band structure, Fermi surface topology and/or scattering from substitutional disorder. It is inherently difficult to disentangle the distinct contributions of each mechanism to a particular property or behavior.
In SDW/SC iron-based superconductors, the effect of disorder has been studied theoretically under the assumption that doping acts as a form of nonmagnetic disorder \cite{fernandes2010,fernandes2012,Vavilov:PRB11,hoyer2014}, and it has received experimental support \cite{Taka2017,KaminskiPRL2011}. The model itself originates from the earlier work by Machida, who studied the competition between CDW and superconductivity in clean systems \cite{Machida}. The key point is that even if nonmagnetic disorder does not directly affect superconductivity, it suppresses CDW long-range order. This suppression, in turn, influences their coupling, e.g., mediated by the phonons, and may lead to a QCP at $T \rightarrow 0$ surrounded by a superconducting ``dome". Moreover, in case of a discontinuous QPT, disorder can change the character of the transition to continuous, enabling (inducing) the QCP \cite{belitz1999,Belitz2005,carlon2001}.

In this article, we demonstrate that controlled point-like disorder can not only suppress CDW but also tune the system to a quantum critical point and beyond. This is shown in a well-known CDW/SC system, Remeika series stannides, $(\text{Ca}_x\text{Sr}_{1-x})_3\text{Rh}_4\text{Sn}_{13}$, by the analysis of temperature-dependent resistivity evolving from Fermi liquid to non-Fermi liquid behavior as a function of added disorder.

Among the Remeika 3-4-13 compounds \footnote{The often-used designation of the 3-4-13 compounds as “quasi-skutterudites” is incorrect from the formal crystallography point of view. Instead, “Remeika series” should be used.}, Ca$_{3}$Rh$_{4}$Sn$_{13}$ has one of the highest superconducting transition temperatures, at $T_c=8.3\:\text{K}$ \cite{Remeika1980}. It does not show a charge-density wave (CDW) or the associated structural distortion. In contrast, its sibling, Sr$_{3}$Rh$_{4}$Sn$_{13}$, develops a second-order structural distortion below $T_{\text{CDW}}=136\:\text{K}$ and superconductivity below $T_{c}=4.6\:\text{K}$. Naturally, the alloy compositions $(\text{Ca}_x\text{Sr}_{1-x})_3\text{Rh}_4\text{Sn}_{13}$, have been investigated for a possible quantum phase transition (QPT), which was found at ambient pressure at the critical concentration, $x_{c}=0.9$, and also by varying isostatic pressure, $P$ \cite{Klintberg2012,Goh2015,Teraski2021}. X-ray measurements confirmed that the $T_{\text{CDW}}\left(x,P\right)$ exists below the superconducting dome \cite{Veiga2020,Carneiro2020, Carneiro2024}. A variety of properties exhibit singular behavior at this structural/CDW quantum critical point (QCP). In addition to signature non-Fermi-liquid behavior from the temperature dependence of the resistivity above $T_{c}$, the specific heat jump at the transition is larger than a weak-coupling value. This is in line with the enhanced superfluid density and pairing strength at the QCP determined by muon spin rotation spectroscopy \cite{Krieger2023}.  The critical current also shows a sharp peak at the QCP composition at temperatures which are deep inside the superconducting dome, further confirming its existence \cite{Liu2022}. In another 'sibling' compound, the refractory metal site is instead permuted, bounded by Ca$_{3}$Rh$_{4}$Sn$_{13}$ (no CDW and $T_{c}=8.3\:\text{K}$) and Ca$_{3}$Ir$_{4}$Sn$_{13}$ ($T_{\text{CDW}}=39\:\text{K}$ and $T_{c}=6.9\:\text{K}$). Recent transport and magnetization measurements indicate that in Ca$_{3}$(Rh$_{x}$Ir$_{1-x}$)$_{4}$Sn$_{13}$ the CDW is suppressed somewhere between $x=0.53$ and $0.58$, and the temperature-dependent resistivity deviates from the Fermi liquid behavior, implying a QCP in this region as well \cite{Krenkel2023}. The interplay between CDW and superconductivity is of active interest in many systems, and a possible connection of CDW fluctuations to superconducting pairing was suggested. These include such diverse systems as kagome lattice materials  \cite{Feng2023,Lin2022,xu2022,Gu2023, Liu2021},  dichalcogenides \cite{Chikina2020}, the cuprates \cite{Miao2021, Jang2017, Lee2014} and the iron pnictides \cite{Bohmer2022}.

To date, the most used nonthermal methods for tuning various systems to a QCP include composition (doping or intercalation), isostatic pressure, (bi)directional strain, and magnetic field. It has been suggested that disorder might also serve as such a control parameter, particularly in cases where the long-range ordered-to-disordered quantum phase transition is not perfectly continuous \cite{Belitz2005}. In these cases, disorder may push the transition toward a pure continuous transition and tune the system toward the QCP \cite{Belitz2005}.
The stability of quantum critical points in disordered systems is typically analyzed using the Harris criterion \cite{harris1974}, which was originally formulated for classical critical points and later extended to quantum random systems \cite{chayes1986}. The criterion stems from the observation that it is not a priori determined whether a disordered system can exhibit a sharp second-order phase transition at a well-defined average value of the coupling parameter, where the system’s response function becomes singular. Since the coupling varies spatially, even if its average value corresponds to criticality, there may exist localized regions where it deviates significantly from the critical point. The Harris condition then arises from the requirement that such strongly non-critical regions are sufficiently rare. This leads to an inequality involving the critical exponent of the correlation length and the spatial dimensionality. However, this approach needs to be conceptually reconsidered when disorder itself serves as a driving force toward criticality. Thus far, only a limited amount of analytical work has been devoted to unraveling the complex phenomena that arise in such situations.

\begin{figure}[tb]
\includegraphics[width=8.4cm]{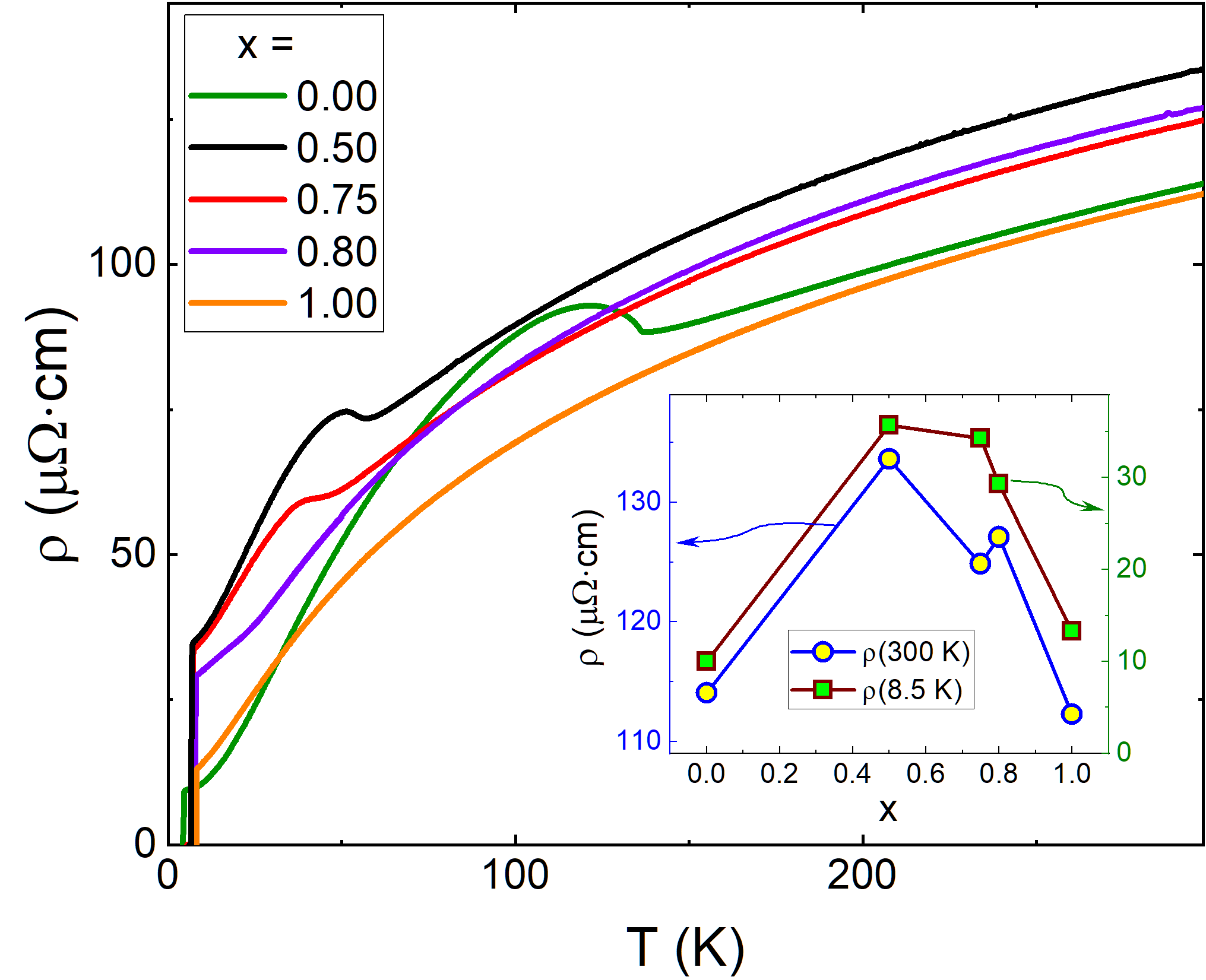}
\caption{(Color online) Temperature-dependent resistivity, $\rho\left(T\right)$,
of pristine samples of $(\text{Ca}_x\text{Sr}_{1-x})_3\text{Rh}_4\text{Sn}_{13}$.
Inset shows compositional evolution of resistivity at two characteristic
temperatures, $\rho\left(T=300\;\text{K}\right)$ (left axis), and
just above the onset of the highest in this set of samples superconducting transition, $\rho\left(8.5\;\text{K}\right)$ (right axis). }
\label{fig:rho(T)}
\end{figure}

Theoretically, suppression of the spin density wave by disorder and appearance of superconductivity has been proposed in the context of the $T_{CDW}(x)$ doping phase diagram of iron-based superconductors \cite{Vavilov:PRB11}. This approach predicts non-Fermi-liquid anomalous behavior in the superconducting \cite{Levchenko2013} and normal state properties, similar to the usual QCP scenario \cite{Custers}, and can lead to the enhancement of superconducting $T_c$ by disorder \cite{FernandesChubukov2012}.

To our knowledge, there have been no reports of a controlled disorder-tuned QCP in any material or for any type of long-range order. Doping influences the chemical potential level via charge doping \cite{Joshi2020} and the electronic band structure via chemical pressure in case of isovalent substitution \cite{Hashimoto2012}. However, it also introduces additional scattering on randomly distributed dopant ions \cite{Joshi2020}. It is impossible to separate these effects and isolate only the effects of scattering or the electronic bandstructure changes.

In this paper, we investigate the effects of disorder on coexisting CDW and superconductivity and demonstrate that the QCP can indeed be reached by introducing a non-magnetic point-like disorder through 2.5 MeV electron irradiation in (Ca$_x$Sr$_{1-x}$)$_3$Rh$_4$Sn$_{13}$ quasi-skutterudite compounds starting with compositions well below the ``compositional'' QCP located at $x_{c}=0.9$.

\section*{Results}

Figure \ref{fig:rho(T)} shows the temperature-dependent resistivity, $\rho\left(T\right)$, of pristine samples of several compositions of $(\text{Ca}_x\text{Sr}_{1-x})_3\text{Rh}_4\text{Sn}_{13}$, indicated in the legend. The inset shows the compositional evolution of $\rho\left(T\right)$ at two characteristic temperatures, $\rho\left(T=300\;\text{K}\right)$ (left $y-$axis), and at the onset of the superconducting transition, $\rho\left(8.5\;\text{K}\right)$ (right $y-$axis).  8.5 K was chosen because it is just above T$_c$ for Ca$_3$Rh$_4$Sn$_{13}$, which is the highest of the compounds shown, allowing for direct comparison between them. To determine the absolute value of the resistivity, we use the geometric factor generated by matching the slope of each curve at high temperature to the known slope of the resistivity of Ca$_3$Rh$_4$Sn$_{13}$ \cite{Krenkel2022}.

\begin{figure}[tb]
\includegraphics[width=7.0cm]{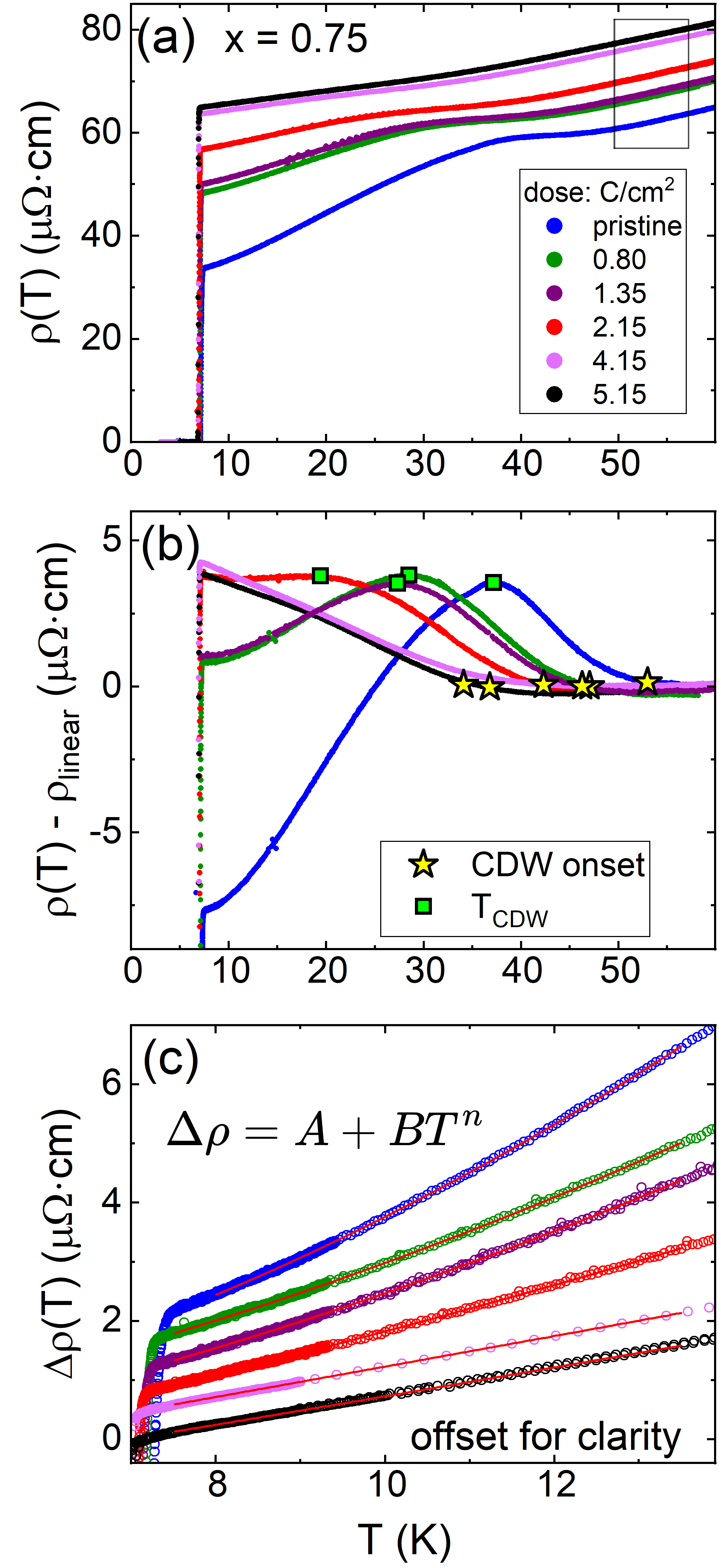}
\caption{(Color online) (a) Temperature-dependent resistivity, $\rho (T)$, for the (Ca$_{0.75}$Sr$_{0.25}$)$_3$Rh$_4$Sn$_{13}$ composition, measured after repeated electron irradiations. The legend shows the accumulated irradiation dose. (b) Determination of the charge density wave onset (yellow stars) and long-range ordering (green stars). The curves were obtained by the subtraction of a linear fit of $\rho (T)$ above the transition temperature in the range indicated by the box in panel (a). (c) Power-law fits, $\Delta\rho=A+B T^n$ of the temperature-dependent resistivity used later for the analysis of the exponent $n$. Here $\Delta\rho=\rho(T)-\rho(8.5\;\text{K})$ and then offset vertically for visual clarity.}
\label{fig:rho(T)diffDose}
\end{figure}

Direct measurements of the geometric factor involve significant errors due to the finite size of the contact spots and non-ideal sample shape. Therefore, to compare different samples and compositions, we assume that at room temperature the dominant contribution to resistivity comes from phonons and that the slope of the curves is similar between the compositions used. This assumption is supported by our earlier measurements on parent compounds \cite{Krenkel2022}. Importantly, at these temperatures, the disorder introduced by electron irradiation leads to a parallel shift of the $\rho(T)$ curves while the slope remains unchanged. Normalization by the slope at high temperatures is a common practice and provides a consistent baseline to measure deviations in the resistivity across different compositions. Additionally, the uncertainty in the absolute resistivity across different compositions does not affect the ratio between the linear and quadratic terms of low-temperature $\rho(T)$.

As expected, the stoichiometric compositions, $x=0$ and $x=1$ show the lowest resistivity values, whereas in mixed compositions the resistivity is higher due to additional substitutional disorder. This makes the initial starting scattering rate of the pristine state $x-$dependent, hence not well defined. Therefore, we rely on the changes in the measured properties introduced by repeated electron irradiation of the same samples.

\begin{figure*}[tb]
\includegraphics[width=14cm]{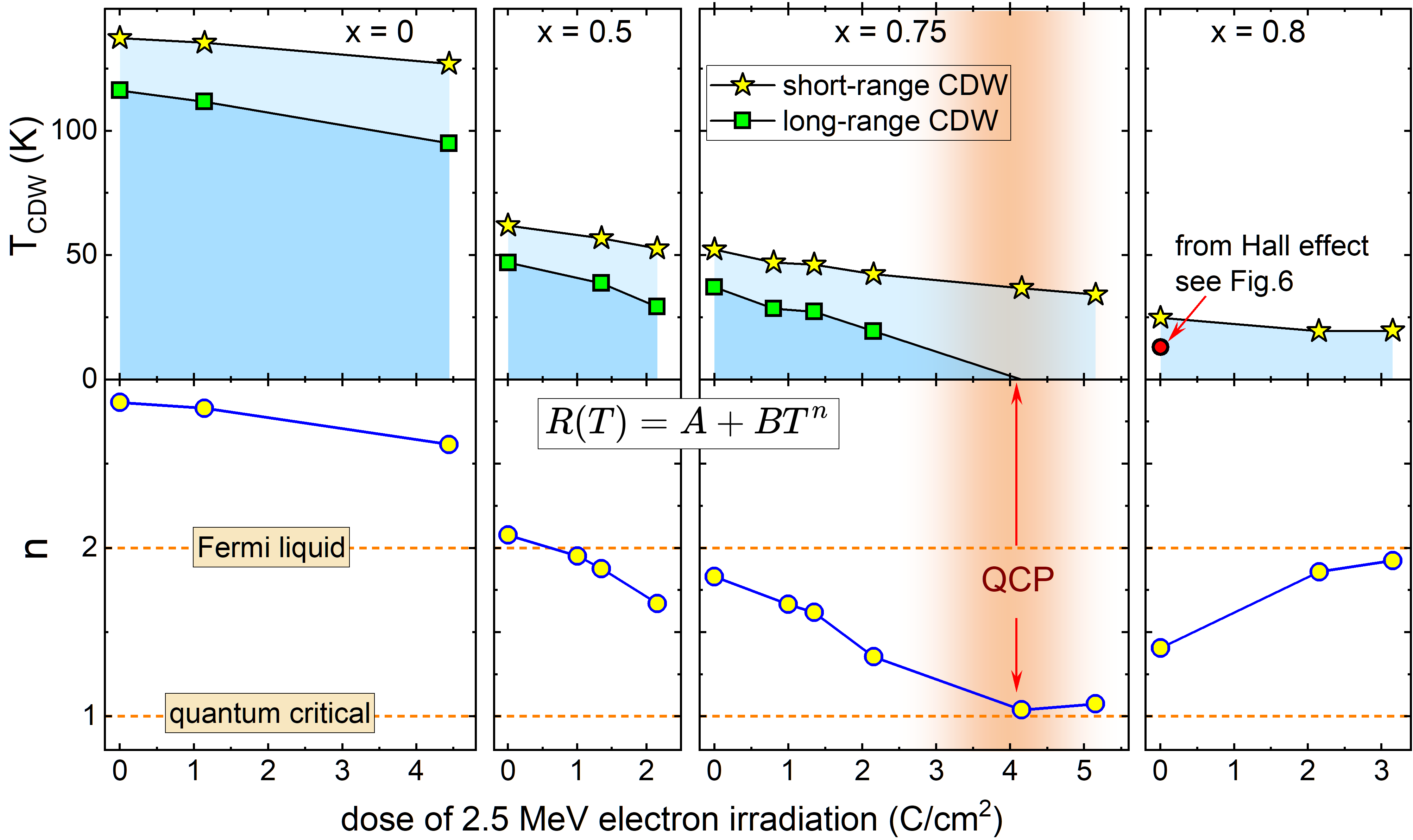}
\caption{(Color online) (Top panels) The dose dependence of the charge density temperatures, $T_{\text{CDW}}\left(x\right)$, in $(\text{Ca}_x\text{Sr}_{1-x})_3\text{Rh}_4\text{Sn}_{13}$ with point-like disorder introduced by electron irradiation. Yellow stars correspond to the onset of deviation from high-temperature $T-$linear behavior in panel (b) of Fig.~\ref{fig:rho(T)diffDose}, indicating the onset of CDW correlations or short-range order. Green stars correspond to a maximum in the difference plot of Fig.~\ref{fig:rho(T)diffDose}, signaling onset of long-range CDW ordering. Bottom panels show evolution of the exponent of the power law fit, as exemplified in the right panel of Fig.~\ref{fig:rho(T)diffDose}.}
\label{fig:TcCDW-Dose}
\end{figure*}

The sample with $x=0.75$, not far from the suggested QCP composition at $x_{c}=0.9$ \cite{Goh2015}, is the main focus of this study, and it was irradiated/measured five times. Figure \ref{fig:rho(T)diffDose}(a) shows the evolution of temperature-dependent resistivity, $\rho\left(T\right)$, with irradiation dose in this sample. Figure \ref{fig:rho(T)diffDose}(b) shows our determination of the onset and the establishment of the long-range CDW order. The curves shown were obtained by the subtraction of a linear fit of $\rho (T)$ above the transition temperature in the range indicated by the box in Fig.\ref{fig:rho(T)diffDose}(a). When the resistivity starts to increase, first CDW correlations appear (marked by yellow stars in Fig.\ref{fig:rho(T)diffDose}(b)), and we associate the peak with the establishment of the long-range CDW order, green stars in Fig.\ref{fig:rho(T)diffDose}(b). This definition is supported by the x-ray diffraction measurements, at least in compositions closer to the QCP, where it was found that for $x$ as compositions as low as $x=0.6$, the long-range CDW order parameter could not be reliably resolved, but there were still well-defined short-range CDW correlations in the material. Those correlations grow stronger as temperature decreases \cite{Upreti2022}. The suppression of CDW by disorder is evident. Not only does the transition temperature shift, but the transition itself broadens. Similar behavior is observed in previously studied stoichiometric Sr$_{3}$Rh$_{4}$Sn$_{13}$ and Sr$_{3}$Ir$_{4}$Sn$_{13}$ \cite{Krenkel2022}. In alloyed compositions, the CDW features are significantly less sharp compared to stoichiometric ones \cite{Krenkel2023}. The resulting CDW temperatures are shown in the top panel of Fig.\ref{fig:TcCDW-Dose} for different irradiation doses, and for samples of the compositions indicated in each panel. The top set of panels shows the evolution of the CDW transition, with a dark shaded area marking the domain of long-range ordering and a light shaded area corresponding to the development of CDW correlations.

Figure~\ref{fig:rho(T)diffDose}(c) shows the fit of the temperature-dependent resistivity curves (offset vertically for visual clarity) to a power law function, $\Delta \rho(T)=A+BT^n$. The evolution of the power law exponent $n$ with disorder in different compositions is summarized in the bottom row of panels in  Fig.\ref{fig:TcCDW-Dose}.

\begin{figure}[tb]
\includegraphics[width=8.6cm]{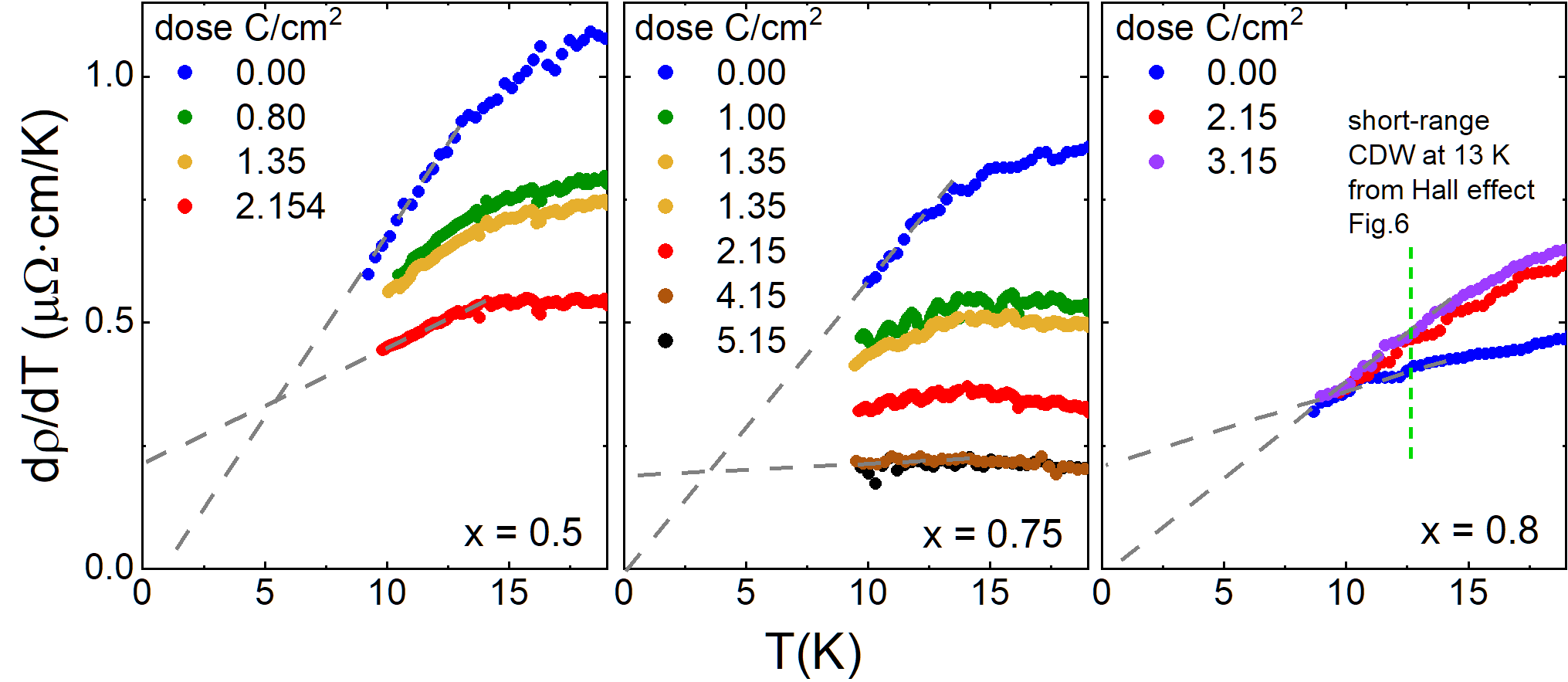}
\caption{(Color online) Derivatives of the resistivity as a function of temperature, for each composition and dose of irradiation indicated in the legends. Vertical line in the right panel shows the onset of CDW transition temperature found in Hall effect measurements Fig.~\ref{fig:Hall} below in pristine sample with $x=0.8$.}
\label{fig:Derivatives}
\end{figure}

The analysis of the transition temperatures suggests that for $x=0.75$ the long-range  $T_{\text{CDW}}$ extrapolates to zero for a dose of 4.1 C/cm$^2$, indicating the possible position of QCP. Interestingly, the analysis of the temperature-dependent resistivity, bottom panels of Fig.~\ref{fig:TcCDW-Dose}, suggests that the minimum exponent, $n$, of the temperature-dependent resistivity reaches $n=1$ at this point, in line with expectations of the quantum critical point scenario. This dose and the region are marked on the figure.

To further justify the power-law analysis, Fig.~\ref{fig:Derivatives} shows the derivative of the resistivity as a function of temperature for four different compositions and varying doses of irradiation. Assuming polynomial functional dependence, $\rho(T)=A_1T+A_2T^2$, we expect $d\rho/dT=A_1+A_2T$, which suggests temperature independent derivative extrapolating to finite value for $T-$linear contribution and $T-$linear derivative for $T^2$ term. As  we approach the quantum critical point, the anomalous $T-$linear term in the resistivity becomes dominant, and we expect the coefficient $A_2$ to decrease and the curve to flatten.  This is precisely the behavior shown in Fig.~\ref{fig:Derivatives}, particularly evident for brown (4.15 C/cm$^2$) and black (5.15 C/cm$^2$) curves in the middle panel. Interestingly, red (2.15 C/cm$^2$) and violet (3.15 C/cm$^2$) curves in the right panel for sample with $x=0.8$ develop $T-$linear derivative extrapolating to zero, revealing emergence of the Fermi-liquid resistivity, $\sim T^2$, clearly in correspondence with power-law analysis in the bottom right panel of  Fig.\ref{fig:TcCDW-Dose}.

We note that in the original research that suggested the QCP in this system, the critical composition was assigned to $x_c=0.9$ from the linear extrapolation of the $T_{CDW}(x)$ to zero \cite{Goh2015}. However, the last composition in which CDW could be resolved from resistivity was $x=0.75$, - the same as in this work. The next composition measured in Ref. \cite{Goh2015} was $x=0.9$, in which CDW had not been detected, and $\rho(T)$ was close to $T-$linear. It is known at least in SDW/SC systems that the transition line can reverse the direction toward smaller $x$ values under the dome of superconductivity \cite{nandi2010,Joshi2020}. Since in this work we have the $x=0.8$ composition where $\rho(T) \approx T^{1.35}$ in the pristine state, we place the QCP in the interval between 0.75 and 0.85.

\begin{figure}[tb]
\includegraphics[width=8.6cm]{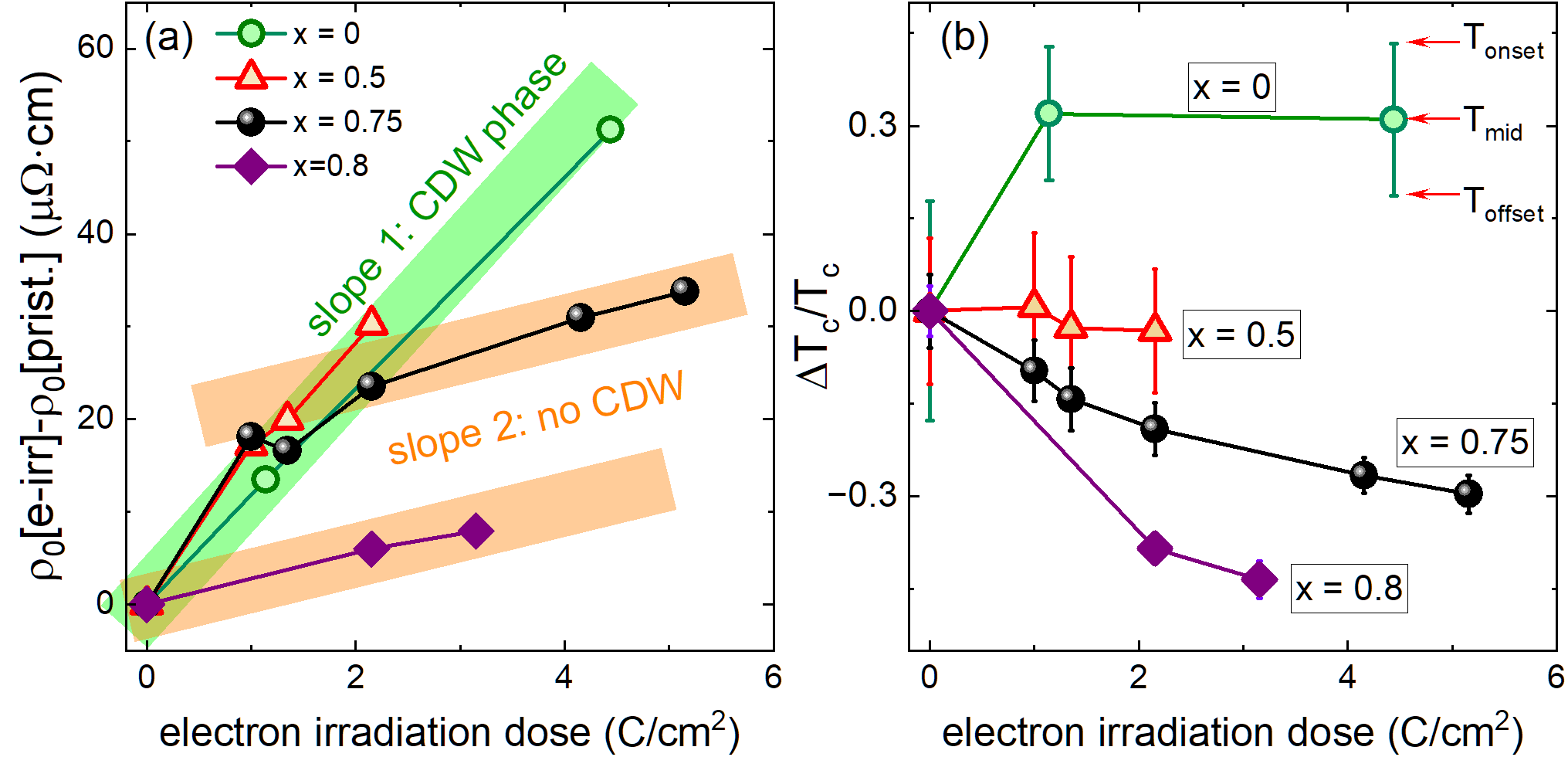}
\caption{(Color online) (a) evolution of the residual resistivity with irradiation dose in compositions as indicated. Note that the slope decreases drastically above $T_{CDW}$, because competing effect of CDW is removed and only disorder scattering remains. (b) evolution of the superconducting $T_c$, presented using a normalized scale, $\Delta T_c/T_c$, for each composition indicated on the legends.}
\label{fig:rho0Tcdose}
\end{figure}

Suppression of the long-range CDW changes the carrier density due to the closing of the partial gap on the Fermi surface. Figure~\ref{fig:rho0Tcdose}(a) shows the evolution of the residual resistivity with irradiation dose. As is clear from the figure, the slope is approximately the same for $x=0$ and $x=0.5$, but decreases dramatically for $x=0.8$. The dependence for $x=0.75$ is non-linear; it initially follows the same slope as $x=$0 but changes to a slope similar to $x=0.8$, providing further evidence for complete CDW suppression for the two highest doses.

Another expectation in the case where quantum fluctuations influence superconductivity is a maximum of $T_c$ in the vicinity of QCP, so that moving towards the QCP increases T$_c$ and moving away from it decreases T$_c$. With disorder as a tuning parameter, the situation becomes more complicated. The right panel of Figure~\ref{fig:rho0Tcdose} shows dose dependence of the superconducting $T_c$, presented using a normalized scale $\Delta T_{c}/T_{c0}\equiv \frac{T_{c}-T_{c0}}{T_{c0}}$, where $T_{c0}$ is the transition temperature before irradiation. Interestingly, the stoichiometric Sr$_{3}$Rh$_{4}$Sn$_{13}$ shows an initial \textit{increase} of $T_{c}$ proving the intrinsic interplay between the CDW and superconductivity. In the simplest scenario, the CDW gapping of the Fermi surface is weakened by the disorder, thereby enhancing the density of states and, therefore, superconductivity. Other mechanisms such as softening of the phonon modes are also possible, as observed in another CDW compound, NbSe$_2$ \cite{KyuilNbSe2}. For the other compositions, the superconducting transition is suppressed, which is consistent with our previous conclusion of the unconventional pairing in Remeika stannide compounds \cite{Krenkel2022}.  However, the rate of change of $T_{c}$ in alloys is difficult to compare directly with the change of $T_c$ in the stoichiometric material. As mentioned above, there is elevated substitutional disorder even in the pristine state of the intermediate compositions, see Fig.\ref{fig:rho(T)}. Since in the absence of nodes, which is the case here, $T_{c}\left(\Gamma\right)$ is a saturating function of the dimensionless scattering rate, $\Gamma$, the rate of change, $dT_{c}\left(\Gamma\right)/d\Gamma$, is also a decreasing function of $\Gamma$.

\begin{figure}[tb]
\includegraphics[width=8cm]{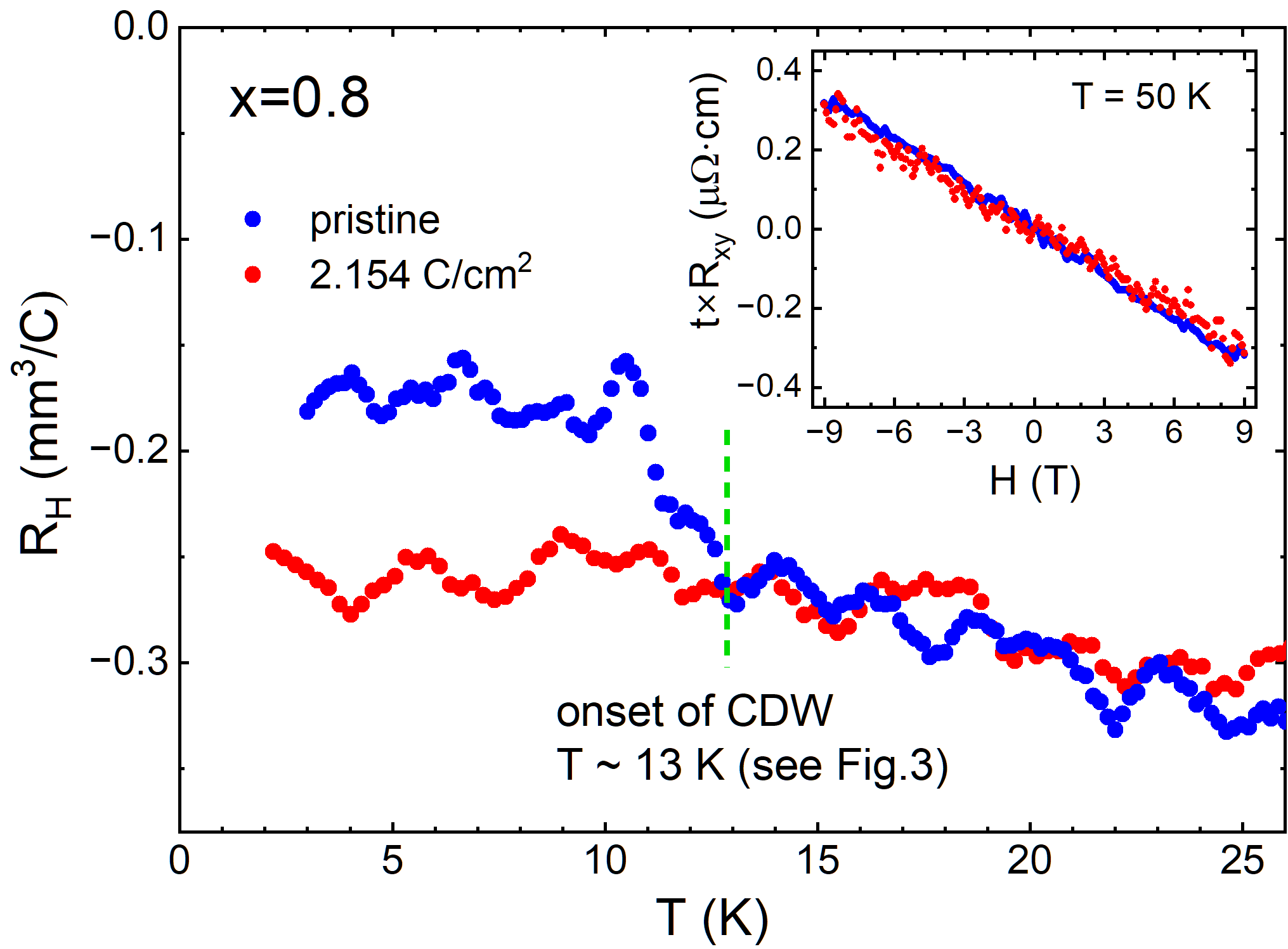}
\caption{(Color online) Temperature dependence of the Hall coefficient before and after irradiation for $x=0.8$ sample. The measurements were taken in constant fields of +/- 9 T as justified by the linear magnetic field variation of the Hall resistance, shown in the inset at $T = 50\;\text{K}$, before and after irradiation.}
\label{fig:Hall}
\end{figure}

To rule out the possibility that the changes that we see in the resistivity are due to changes in the carrier concentration induced by electron irradiation, we also measured the Hall effect before and after irradiation in the $x = 0.8$ compound. The results are shown in Fig.~\ref{fig:Hall}. At $50\;\text{K}$ (well above the CDW transition), the Hall resistance is a linear function of a magnetic field and there is no discernible difference before and after irradiation. A similar result has been obtained in other stannides and SDW/SC systems \cite{npjProzorov2019}.
Therefore, we are confident that the changes in resistivity with disorder are not due to charge ``doping'' from electron irradiation. Furthermore, it is established that in 3-4-13 compounds, the Hall effect is temperature dependent and shows clear anomalies at the $T_{\text{CDW}}$, for example, in Sr$_3$Ir$_4$Sn$_{13}$ and Ca$_3$Ir$_4$Sn$_{13}$ \cite{Tseng2018,Wang2015}. Figure~\ref{fig:Hall} shows that in our case, the Hall constant remains unchanged by irradiation down to 13 K, at which the pristine sample shows an anomaly associated with the onset of CDW transition. Suppression of this feature after irradiation further supports the conclusion that disorder drives the CDW phase transition to zero.

\section*{Discussion}

As noted in the introduction, several phenomena are expected near a quantum critical point. In particular, a non-Fermi liquid behavior is observed not only at the QCP but also well above and around it. Experimentally, this is inferred from the observed suppression of CDW temperature by disorder and simultaneous emergence of a $T-$linear resistivity in a $x=0.75$ sample, Fig.~\ref{fig:TcCDW-Dose}. Moving further away from the QCP, a Fermi-liquid state emerges, as evidenced by the reappearance of a $T^2-$ resistivity. This is indeed observed from both power-law behavior and derivative analysis. Suppressing the short-range leftover CDW in $x=0.8$ sample enhances a $T^2$ term, suggesting that we not only tuned the system to the QCP but likely have overshot it and moved deeper into the Fermi-liquid regime.

The beauty of quantum criticality is that the behavior around QCP is independent of a particular mechanism of how it was reached as long as it stays in the same universality class described by a set of critical exponents. Therefore, the particular mechanism by which disorder drives the system is the one that suppresses the CDW order. This is an active area of research and there are several mechanisms. There is a general statement known as Imry-Ma argument stating that for a continuous symmetry, random fields (like those introduced by disorder) destabilize long-range order in dimensions $D \leq 4$,\cite{imry1975,changlani2016}. Another often discussed specific mechanism of CDW suppression is phase pinning that disrupts global CDW coherence, leading to domains of short-range order, described by the Fukuyama–Lee–Rice model \cite{imry1975,fukuyama1978}.
Furthermore, inhomogeneous local strain and electronic structure alterations may disrupt electron-phonon interaction, affecting specific phonon modes associated with the CDW wavevector; by broadening the electronic states, disorder weakens the nesting conditions. Also, Friedel oscillations around defects have a wavevector related to the Fermi momentum of the material \cite{dobson1995}. These localized oscillations can interfere with the global charge modulation of the CDW, further contributing to the disruption of its long-range order.

\section*{Conclusions}

To summarize, a controlled non-magnetic point-like disorder has been successfully used as a novel tuning parameter to suppress the CDW and reach the QCP under the dome of superconductivity in single crystals of $(\text{Ca}_x\text{Sr}_{1-x})_3\text{Rh}_4\text{Sn}_{13}$. Starting at compositions below the suggested structural QCP, disorder induced by 2.5 MeV electron irradiation resulted in a progressively larger linear term and a reduced quadratic term in resistivity, $\rho\left(T\right)$, just above $T_c$. The behavior becomes almost perfectly linear at 4.14 C/cm$^2$ and starts moving toward quadratic, suggesting that disorder pushes the system even beyond the QCP. These results strongly support the general prediction that the non-magnetic disorder can tune the system toward the quantum critical regime even in the case of two coexisting long-range quantum orders.

The ability to tune charge-density wave (CDW)/superconductivity systems to a quantum critical point (QCP) using controlled disorder opens new avenues for exploring quantum criticality and associated phenomena, such as possible tuning of the tricritical point to a QCP,  Griffiths singularities in rare regions, as well as another way to explore unconventional superconductivity \cite{belitz2005a,Vojta2006,liu2024d}. This approach complements traditional tuning parameters and provides a framework for future studies in other correlated electron systems, such as high-temperature superconductors, heavy-fermion, and topologically nontrivial materials. The findings reported here underscore the potential of controlled disorder as a non-thermal tuning parameter to uncover novel quantum states, with implications for both fundamental research and potential technological applications.

\section*{Materials and Methods}

\textbf{Samples:} Single crystals $(\text{Ca}_x\text{Sr}_{1-x})_3\text{Rh}_4\text{Sn}_{13}$ were
grown from tin flux \cite{Krieger2023}. A mixture of 3 parts (Ca + Sr of the desired ratio), 4 parts Rh and 93 parts Sn, was placed in an alumina crucible and sealed under vacuum in a quartz ampule along with a bit of quartz wool for filtration. The ampule was heated up to 1100~$^o$C, kept for six hours, fast cooled to 800~$^o$C and then cooled slowly to 490 $^o$C where the excess Sn was decanted in a centrifuge outside the furnace. The process is described elsewhere \cite{cairsn1}. The obtained samples were characterized by powder X-ray diffraction and energy-dispersive X-ray compositional analysis. This well-studied system shows good sample homogeneity and reproducibility of resistivity and EDX compositions between samples.

\textbf{Transport measurements:} Electrical resistivity was measured in a standard four-probe configuration in bar-shaped single crystals of typical length of 1-3 mm, width of 0.3-0.7 mm and thickness between 50 and 300 $\mu$m. To prepare the samples, the crystals were etched with HCl, cut to dimensions using a wire saw, and polished. The electrical contacts were formed by soldering 50 $\mu$m silver wires with tin-silver solder \cite{Krenkel2022,Krenkel2023} with typical contact resistances below 100 $\mu\Omega$.
\textbf{Electron irradiation:} The 2.5 MeV electron irradiation was performed at the ``SIRIUS'' accelerator facility at Laboratoire des Solides Irradi\'{e}s, \'{E}cole Polytechnique, Palaiseau, France. The acquired irradiation dose is measured by a calibrated Faraday trap behind the sample and is conveniently expressed in $\textrm{C}/\textrm{cm}^{2}$, where 1 $\textrm{C}/\textrm{cm}^{2}=6.24\times10^{18}\:\textrm{e}^{-}/\textrm{cm}^{2}$. Collisions of electrons with ions produce Frenkel pairs (vacancy + interstitial) because the relativistic energy transfer upon collisions matches well with the knockout energy threshold, typically between 20 and 80 eV. At these energies, the electron scattering cross-section is on the order of 100 barns. Other particles used for irradiation have either much lower cross-sections (neutrons and gamma-rays), or may produce clusters (protons), and other correlated defects, including columnar tracks (heavy ions).

Another important parameter for irradiation is the particles' penetration depth into the material. While heavier particles require very thin samples because of beam attenuation, electrons produce relatively uniform disorder at mm-range depths for typical samples. Importantly, for the doses of irradiation studied, the produced defects are in the dilute limit, typically with about one defect per thousands of host lattice ions. Furthermore, according to extensive Hall-effect measurements, the electron irradiation does not ``charge-dope'' the samples and does not cause a shift of the chemical potential \cite{npjProzorov2019,ProzorovPhysRevX}.

The irradiation is conducted at low temperatures with the sample immersed in liquid hydrogen around 22 K. This is done to remove heat generated upon collisions and to prevent immediate recombination and clustering of the defects. Upon warming to room temperature, some pairs recombine, and some defects migrate to various sinks such as lattice defects, dislocations, and surfaces, leaving a metastable population of point-like defects behind, typically about 70\% of the initially produced population. The final amount of the induced disorder is directly estimated from the changes in residual resistivity. Importantly, to avoid variation between different samples, in our experiments, the same sample was repeatedly irradiated and then measured to observe the progression of the results. Several compositions around the QCP concentration were studied. For a more detailed discussion of the use of controlled electron irradiation to study materials, the reader is referred to specialized books \cite{Damask1963,Thompson1969}.

\section*{Acknowledgements}

This work was supported by the National Science Foundation under Grant No. DMR-2219901. M.A.T. and K.R.J. were supported by the U.S. Department of Energy (DOE), Office of Science, Basic Energy Sciences, Materials Science and Engineering Division. Ames National Laboratory is operated for the U.S. DOE by Iowa State University under Contract No. DE-AC02-07CH11358. A.L. acknowledges the financial support by the National Science Foundation Grant No. DMR-2203411 and a Research Fellowship funded by the Alexander von Humboldt Foundation. Work in France was supported by “Investissements d’Avenir” LabEx PALM (Grant No. ANR10-LABX-0039-PALM). The authors acknowledge support from the EMIR\&A French network (FR CNRS 3618) on the platform SIRIUS, proposals No. 20-5925 and 23-4663. Work at BNL (materials synthesis) was supported by the U.S. Department of Energy, Basic Energy Sciences, Division of Materials Science and Engineering, under Contract No. DE-SC0012704. C.P. acknowledges support from the Shanghai Key Laboratory of Material Frontiers Research in Extreme Environments, China (No. 22dz2260800) and Shanghai Science and Technology Committee, China (No. 22JC1410300).

\bibliographystyle{apsrev4-2.bst}
%

\end{document}